# Classical-field description of the quantum effects in the light-atom interaction

**Sergey A. Rashkovskiy**

*Institute for Problems in Mechanics of the Russian Academy of Sciences, Vernadskogo Ave., 101/1, Moscow, 119526, Russia*

*Tomsk State University, 36 Lenina Avenue, Tomsk, 634050, Russia*

*E-mail: rash@ipmnet.ru, Tel. +7 906 0318854*

March 1, 2016

**Abstract**  In this paper I show that light-atom interaction can be described using purely classical field theory without any quantization. In particular, atom excitation by light that accounts for damping due to spontaneous emission is fully described in the framework of classical field theory. I show that three well-known laws of the photoelectric effect can also be derived and that all of its basic properties can be described within classical field theory.

## 1 Introduction

The atom-field interaction is one of the most fundamental problems of quantum optics. It is believed that a complete description of the light-atom interaction is possible only within the framework of quantum electrodynamics (QED), when not only the states of an atom are quantized but also the radiation itself. In quantum mechanics, this viewpoint fully applies to iconic phenomena, such as the photoelectric effect, which cannot be explained in terms of classical mechanics and classical electrodynamics according to the prevailing considerations. There were attempts to build a so-called semiclassical theory in which only the states of an atom are quantized while the radiation is considered to be a classical Maxwell field [1-6]. Despite the success of the semiclassical approach, there are many optical phenomena that cannot be described by the semiclassical theory. It is believed that an accurate description of these phenomena requires a full quantum-mechanical treatment of both the atom and the field. In addition, it is clear that this point of view is inconsistent when we abandon the quantization of the radiation, from which, in fact, quantum mechanics started, but continue to quantize the atom. In previous papers of this series [7-11], an attempt was made to construct a completely classical theory, which is similar to classical field theory [12], in which any quanta are absent. Thus, in papers [7-9], it was shown that the discrete events (e.g., clicks of a detector, emergence of the spots on a photographic plate) that are observed in some of the "quantum" experiments with light (especially in the double-slit experiments), which are considered to be direct evidence of the existence of photons, can in fact be explained within classical electrodynamics without quantization of the radiation. Similarly, if the electrons are considered to not be a particle but instead a classical continuous wave field, similar to the classical electromagnetic field, one can consistently explain the "wave-particle duality of electrons" in the double-slit experiments [10]. In this case, the Dirac equation and its specific cases (Klein-Gordon, Pauli and Schrödinger) should be considered to be the usual field equations of a classical electron wave field, similar to Maxwell's equations for classical electromagnetic fields. As was shown in [10], considering the electron wave as a classical field, we must assign to it, besides the energy and momentum which are distributed in space, also an electric charge, an internal angular momentum and an internal magnetic moment, which are also continuously distributed in space. In this case, the internal angular momentum and internal magnetic moment of the electron wave are its intrinsic properties and cannot be reduced to any movement of charged particles. This viewpoint allows for a description in natural way, in the framework of classical field theory with respect to the many observed phenomena that involve "electrons", and it explains their properties which are considered to be paradoxical from the standpoint of classical mechanics. Thus, the Compton Effect, which is considered to be "direct evidence of the existence of photons", has a natural explanation if both light and electron waves are considered to be classical continuous fields [10]. The same approach can be applied to the Born rule for light and "electrons" as well as to the Heisenberg uncertainty principle, which have a simple and clear explanation within classical field theory [7-10]. Using such a point of view on the nature of the "electron", a new model of the hydrogen atom that differs from the conventional planetary model was proposed and justified in [11]. According to this model, the atom represents a classical open volume resonator in which an electrically charged continuous electron wave is held in a restricted region of space by the electrostatic field of the nucleus. As shown in [11], the electrostatic field of the nucleus plays for the electron wave the role of a "dielectric medium", and thus, one can say that the electron wave is held in the hydrogen atom due to the total internal reflection on the inhomogeneities of this "medium". In the hydrogen atom, as in any volume resonator, there are eigenmodes that correspond to a discrete spectrum of eigenfrequencies, which are the eigenvalues of the field equation (e.g., Schrödinger, Dirac). As usual, the standing waves (in this case, the standing electron waves) correspond to the eigenmodes. If only one of the eigenmodes is excited in the atom as in the volume resonator, then such a state of the atom is called a pure state. If simultaneously several (two or more) eigenmodes are excited in the atom, then such a state is called a mixed state [11].

Using this viewpoint, it was shown in [11] that all of the basic optical properties of the hydrogen atom have a simple and clear explanation in the framework of classical electrodynamics without any quantization. In particular, it was shown that the atom can be in a pure state indefinitely. This arrangement means that the atom has a discrete set of stationary states, which correspond to all possible pure states, but only the pure state that corresponds to the lowest eigenfrequency is stable. Precisely this state is the ground state of the atom. The remaining pure states are unstable, although they are the stationary states. Any mixed state of an atom in which several eigenmodes are excited simultaneously is non-stationary, and according to classical electrodynamics, the atom that is in that state continuously emits electromagnetic waves of the discrete spectrum, which is interpreted as a spontaneous emission.

In reference [11], a fully classical description of spontaneous emission was given, and all of its basic properties that are traditionally described within the framework of quantum electrodynamics were obtained. It is shown that the "jump-like quantum transitions between the discrete energy levels of the atom" do not exist, and the spontaneous emission of an atom occurs not in the form of discrete quanta but continuously.

As is well known, the linear wave equation, e.g., the Schrödinger equation, cannot explain the spontaneous emission and the changes that occur in the atom in the process of spontaneous emission (so-called "quantum transitions"). To explain spontaneous transitions, quantum mechanics must be extended to quantum electrodynamics, which introduces such an object as a QED vacuum, the fluctuations of which are considered to be the cause of the "quantum transitions".

In reference [11], it was shown that the Schrödinger equation, which describes the electron wave as a classical field, is sufficient for a description of the spontaneous emission of a hydrogen atom. However, it should be complemented by a term that accounts for the inverse action of self-electromagnetic radiation on the electron wave. In the framework of classical electrodynamics, it was shown that the electron wave as a classical field is described in the hydrogen atom by a nonlinear equation [11]

$$i\hbar \frac{\partial \psi}{\partial t} = -\frac{1}{2m_e}\Delta\psi - \frac{e^2}{r}\psi - \frac{2e^2}{3c^3}\psi \mathbf{r}\frac{\partial^3}{\partial t^3}\int \mathbf{r}|\psi|^2 d\mathbf{r} \qquad (1)$$

where the last term on the right-hand side describes the inverse action of the self-electromagnetic radiation on the electron wave and is responsible for the degeneration of any mixed state of the hydrogen atom. Precisely, this term "provides" a degeneration of the mixed state of the hydrogen atom to a pure state, which corresponds to the lower excited eigenmodes of an atom. As shown in [11], this term has a fully classical meaning and fits into the concept developed in [7-11] in that the photons and electrons as particles do not exist, and there are only electromagnetic and electron waves, which are classical (continuous) fields.

In this paper, using equation (1), the light-atom interactions are considered from the standpoint of classical field theory. In particular, two effects will be considered: (i) an excitation of the mixed state of the atom by an electromagnetic wave without ionization of the atom; and (ii) the photoelectric effect, in which the action of the electromagnetic wave on the atom causes its ionization.

## 2 Optical equations with damping due to spontaneous emission

Let us consider a hydrogen atom that is in a light field, while accounting for its spontaneous emission.

Under the action of the light field, the excitation of some eigenmodes of the electron wave in the hydrogen atom at the expense of others occurs. This phenomenon will be manifested in a stimulated redistribution of the electric charge of the electron wave in the atom between its eigenmodes that occur under the action of the nonstationary electromagnetic field [11]. As a result, the atom, as an open volume resonator, will go into a mixed state in which, in full compliance with classical electrodynamics, electric dipole radiation will occur [11], which is customarily called a spontaneous emission. As shown in [11], the emission will be accompanied by a spontaneous "crossflow" of the electric charge of the electron wave from an eigenmode that has a greater frequency to an eigenmode that has a lower frequency. Under certain conditions, between the spontaneous and stimulated redistribution of the electric charge of an electron wave within the atom, a detailed equilibrium can be established, in which the amount of electric charge that crossflows from an eigenmode $k$ into an eigenmode $n$ per unit time will be equal to the amount of electric charge that crossflows from eigenmode $n$ into eigenmode $k$ per unit time. Clearly, such an equilibrium will be forced and will depend on the intensity of the light field. This is easily seen when the light field is instantaneously removed. As shown in [11], the atom, which was previously in a mixed state, will spontaneously emit electromagnetic waves, and this spontaneous emission will be accompanied by a spontaneous redistribution of the electric charge of the electron wave between the excited eigenmodes of the atom. As a result, over time, the whole electric charge of the electron wave "crossflows" into a lower eigenmode (i.e., with a lower frequency) of the initially excited eigenmodes and the atom enters into a pure state which can be retained indefinitely because spontaneous emission is absent in this state.

Let us consider this process in more detail by the example of the hydrogen atom.

If an atom is in an external electromagnetic field, then equation (1) should be supplemented by the terms which take into account the external action. As a result, one obtains the equation

$$i\hbar \frac{\partial \psi}{\partial t} = \left[\frac{1}{2m_e}\left(\frac{\hbar}{i}\nabla + \frac{e}{c}\mathbf{A}\right)^2 - e\left(\frac{e}{r} + \varphi\right) - \frac{2e^2}{3c^3}\mathbf{r}\frac{\partial^3}{\partial t^3}\int \mathbf{r}|\psi|^2 d\mathbf{r}\right]\psi \quad (2)$$

where $\varphi(t, \mathbf{r})$ and $\mathbf{A}(t, \mathbf{r})$ are the scalar and vector potentials of an external electromagnetic field.

We consider a linearly polarized electromagnetic wave with the wavelength $\lambda$, which is much more than the characteristic size of the hydrogen atom:

$$\lambda \gg \int r|\psi|^2 d\mathbf{r} \quad (3)$$

In this case, one can consider the electric field of the electromagnetic wave in the vicinity of the hydrogen atom to be homogeneous but nonstationary: $\mathbf{E}(t, \mathbf{r}) = \mathbf{E}_0 \cos \omega t$, where $\mathbf{E}_0$ and $\omega$ are constants. For such a field, one can select the gauge [12], at which

$$\mathbf{A} = 0, \varphi = -\mathbf{r}\mathbf{E}_0 \cos \omega t \quad (4)$$

In this case, equation (2) takes the form

$$i\hbar \frac{\partial \psi}{\partial t} = -\frac{1}{2m_e}\Delta\psi - \frac{e^2}{r}\psi + \psi e\mathbf{r}\mathbf{E}_0 \cos \omega t - \frac{2e^2}{3c^3}\psi \mathbf{r}\frac{\partial^3}{\partial t^3}\int \mathbf{r}|\psi|^2 d\mathbf{r} \quad (5)$$

As usual, the solution of equation (5) can be found in the form

$$\psi(t, \mathbf{r}) = \sum_n c_n(t) u_n(\mathbf{r}) \exp(-i\omega_n t) \quad (6)$$

where the constants $\omega_n$ and the functions $u_n(\mathbf{r})$ are the eigenvalues and eigenfunctions of the linear Schrödinger equation

$$\hbar \omega_n u_n = -\frac{1}{2m_e}\Delta u_n - \frac{e^2}{r}u_n \quad (7)$$

The functions $u_n(\mathbf{r})$ form the orthonormal system:

$$\int u_n(\mathbf{r}) u_k^*(\mathbf{r}) d\mathbf{r} = \delta_{nk} \quad (8)$$

Substituting (6) into (5) while accounting for (7) and (8), we obtain

$$i\hbar \frac{dc_n}{dt} = -\sum_k c_k (\mathbf{d}_{nk}\mathbf{E}_0) \exp(i\omega_{nk}t) \cos \omega t - \frac{2}{3c^3}\sum_k c_k (\mathbf{d}_{nk}\dddot{\mathbf{d}}) \exp(i\omega_{nk}t) \quad (9)$$

where

$$\omega_{nk} = \omega_n - \omega_k \quad (10)$$

$$\mathbf{d} = -e \int \mathbf{r}|\psi|^2 d\mathbf{r} \quad (11)$$

is the electric dipole moment of the electron wave in the hydrogen atom:

$$\mathbf{d}_{nk} = \mathbf{d}_{kn}^* = -e \int \mathbf{r} u_n^*(\mathbf{r}) u_k(\mathbf{r}) d\mathbf{r} \quad (12)$$

Considering expressions (6), (11) and (12), we can write

$$\mathbf{d} = \sum_n \sum_k c_k c_n^* \mathbf{d}_{nk} \exp(i\omega_{nk}t) \quad (13)$$

Let us consider the so-called "two-level atom", i.e., the case in which only two eigenmodes $k$ and $n$ of an atom are excited simultaneously.

For definiteness, one assumes that $\omega_n > \omega_k$, and correspondingly,

$$\omega_{nk} > 0 \tag{14}$$

In this case, equations (9) and (11) take the form

$$i\hbar \frac{dc_n}{dt} = -[c_k(\mathbf{d}_{nk}\mathbf{E}_0)\exp(i\omega_{nk}t) + c_n(\mathbf{d}_{nn}\mathbf{E}_0)]\cos\omega t - \frac{2}{3c^3}[c_k(\mathbf{d}_{nk}\ddot{\mathbf{d}})\exp(i\omega_{nk}t) + c_n(\mathbf{d}_{nn}\ddot{\mathbf{d}})] \tag{15}$$

$$i\hbar \frac{dc_k}{dt} = -[c_n(\mathbf{d}_{nk}^*\mathbf{E}_0)\exp(-i\omega_{nk}t) + c_k(\mathbf{d}_{kk}\mathbf{E}_0)]\cos\omega t - \frac{2}{3c^3}[c_n(\mathbf{d}_{nk}^*\ddot{\mathbf{d}})\exp(-i\omega_{nk}t) + c_k(\mathbf{d}_{kk}\ddot{\mathbf{d}})] \tag{16}$$

$$\mathbf{d} = |c_n|^2\mathbf{d}_{nn} + |c_k|^2\mathbf{d}_{kk} + c_k c_n^*\mathbf{d}_{nk}\exp(i\omega_{nk}t) + c_n c_k^*\mathbf{d}_{nk}^*\exp(-i\omega_{nk}t) \tag{17}$$

In general, differentiating vector $\mathbf{d}$ with respect to time, we must account for the fact that the parameters $c_n$ are functions of time. As shown below, the parameters $c_n$ are changed with time much more slowly than the oscillating factor $\exp(i\omega_{nk}t)$. This arrangement means that there is the condition

$$|\dot{c}_n| \ll \omega_{nk}|c_n| \tag{18}$$

Considering (17) and (18), we obtain approximately

$$\ddot{\mathbf{d}} = i\omega_{nk}^3[c_n c_k^* \mathbf{d}_{nk}^* \exp(-i\omega_{nk}t) - c_k c_n^* \mathbf{d}_{nk}\exp(i\omega_{nk}t)] \tag{19}$$

Substituting expression (19) into equations (15) and (16), we obtain

$$i\hbar \frac{dc_n}{dt} = -[c_k(\mathbf{d}_{nk}\mathbf{E}_0)\exp(i\omega_{nk}t) + c_n(\mathbf{d}_{nn}\mathbf{E}_0)]\cos\omega t - \frac{2}{3c^3}i\omega_{nk}^3[c_n|c_k|^2|\mathbf{d}_{nk}|^2 + c_n c_n c_k^*(\mathbf{d}_{nn}\mathbf{d}_{nk}^*)\exp(-i\omega_{nk}t)] + \frac{2}{3c^3}i\omega_{nk}^3[c_k c_k c_n^*(\mathbf{d}_{nk})^2\exp(2i\omega_{nk}t) + |c_n|^2 c_k(\mathbf{d}_{nn}\mathbf{d}_{nk})\exp(i\omega_{nk}t)] \tag{20}$$

$$i\hbar \frac{dc_k}{dt} = -[c_n(\mathbf{d}_{nk}^*\mathbf{E}_0)\exp(-i\omega_{nk}t) + c_k(\mathbf{d}_{kk}\mathbf{E}_0)]\cos\omega t - \frac{2}{3c^3}i\omega_{nk}^3[c_n c_n c_k^*(\mathbf{d}_{nk}^*)^2\exp(-2i\omega_{nk}t) + |c_k|^2 c_n(\mathbf{d}_{kk}\mathbf{d}_{nk}^*)\exp(-i\omega_{nk}t)] + \frac{2}{3c^3}i\omega_{nk}^3[c_k|c_n|^2|\mathbf{d}_{nk}|^2 + c_k c_k c_n^*(\mathbf{d}_{kk}\mathbf{d}_{nk}\exp(i\omega_{nk}t))] \tag{21}$$

where $|\mathbf{d}_{nk}|^2 = (\mathbf{d}_{nk}\mathbf{d}_{nk}^*)$, $(\mathbf{d}_{nk})^2 = (\mathbf{d}_{nk}\mathbf{d}_{nk})$.

Equations (20) and (21) contain rapidly oscillating terms with frequencies of $\omega$, $\omega_{nk}$ and $2\omega_{nk}$, which in view of (18) can be removed by averaging over the fast oscillations. As a result, we obtain the equations

$$i\hbar \frac{dc_n}{dt} = -c_k(\mathbf{d}_{nk}\mathbf{E}_0)\exp(i\omega_{nk}t)\cos\omega t - i\frac{2\omega_{nk}^3}{3c^3}c_n|c_k|^2|\mathbf{d}_{nk}|^2 \tag{22}$$

$$i\hbar \frac{dc_k}{dt} = -c_n(\mathbf{d}_{nk}^*\mathbf{E}_0)\exp(-i\omega_{nk}t)\cos\omega t + i\frac{2\omega_{nk}^3}{3c^3}c_k|c_n|^2|\mathbf{d}_{nk}|^2 \tag{23}$$

Equations (22) and (23) describe the Rabi oscillations with damping due to spontaneous emission. The first term on the right-hand side of equation (22) describes the excitation of mode $n$ due to the impact of the incident electromagnetic wave on mode $k$. In quantum mechanics, it is traditionally interpreted to be an induced transition from a lower energy level $k$ to a higher energy level $n$ due to "absorption of the photon". The first term on the right-hand side of equation (23) describes the excitation of mode $k$ due to the impact of the incident electromagnetic wave on mode $n$. In quantum mechanics, it is traditionally interpreted as an induced transition from a higher energy level $n$ to a lower energy level $k$ due to the "emission of a photon". The last term on the right-hand side of equations (22) and (23) is traditionally interpreted as a spontaneous transition from a higher level $n$ to a lower level $k$, which is accompanied by an " emission of a photon" with a frequency of $\omega_{nk}$.

From the viewpoint of classical field theory considered here, the first terms on the right-hand sides of equations (22) and (23) describe an induced (by the action of the light field) redistribution of the electric charge of the continuous electron wave between the excited modes $n$ and $k$, while the second terms describe a spontaneous redistribution of the electric charge between these two modes, which is accompanied by a spontaneous emission. In our analysis, there are neither photons nor electrons; there are no "jump-like transitions" between the atom's discrete energy levels, which, incidentally, are also absent, while the terms of equations (22) and (23) describe the interaction of two classical wave fields – the electromagnetic wave and the electron wave.

Let us introduce the notation

$$\rho_{nn} = |c_n|^2, \rho_{kk} = |c_k|^2, \rho_{nk} = c_n c_k^*, \rho_{kn} = c_k c_n^* \tag{24}$$

Obviously,

$$\rho_{nn} + \rho_{kk} = 1 \tag{25}$$

$$\rho_{nk} = \rho_{kn}^* \tag{26}$$

Condition (25) according to [11] expresses the law of conservation of electric charge, which means that in the process under consideration, the electric charge of the electron wave is simply redistributed between modes $n$ and $k$. Note that in this section, we do not consider the photoelectric effect, i.e., an emission of an electron wave by the atom under the influence of an incident electromagnetic wave; the photoelectric effect will be discussed below.

Using equations (22) and (23) for the parameters in (24), we obtain the equation

$$\frac{d\rho_{nn}}{dt} = -\frac{d\rho_{kk}}{dt} = i[\rho_{kn} b_{nk} \exp(i\omega_{nk}t) - \rho_{nk} b_{nk}^* \exp(-i\omega_{nk}t)] \cos \omega t - 2\gamma_{nk}\rho_{nn}\rho_{kk} \tag{27}$$

$$\frac{d\rho_{nk}}{dt} = \frac{d\rho_{kn}^*}{dt} = (\rho_{kk} - \rho_{nn})[ib_{nk} \exp(i\omega_{nk}t) \cos \omega t - \gamma_{nk}\rho_{nk}] \tag{28}$$

where

$$\gamma_{nk} = \frac{2\omega_{nk}^3}{3\hbar c^3} |\mathbf{d}_{nk}|^2 \tag{29}$$

$$b_{nk} = b_{kn}^* = \frac{1}{\hbar}(\mathbf{d}_{nk}\mathbf{E}_0) \tag{30}$$

Equations (27) and (28) describe the interaction of electromagnetic waves with a hydrogen atom while accounting for a spontaneous emission. They, in fact, are *the optical equations with damping due to spontaneous emission* [13]. However, there are fundamental differences in equations (27) and (28) from the conventional (linear) optical equations [13]. Thus, in quantum optics, the optical equations that account for damping due to spontaneous emission are not derived strictly but are postulated by the addition of the corresponding linear damping terms in the linear wave equations. This arrangement is because the damping in a spontaneous emission cannot be obtained from the linear Schrödinger equation and requires a "second quantization". In our approach, equations (27) and (28) are strictly derived and are a direct and natural consequence of the nonlinear Schrödinger equation (5), which accounts for the spontaneous emission. Equations (27) and (28) are nonlinear in contrast to conventional (linear) optical equations because the last terms on the right-hand side of equations (27), (28) non-linearly depend on $\rho_{kk}$ and $\rho_{nn}$. They become almost linear if the excitation of the upper mode $n$ is weak, and we have approximately $\rho_{kk} \approx \rho_{kk} - \rho_{nn} \approx 1$. In this case, equations (27) and (28) turn into the conventional (linear) optical equations [13]. Moreover, the damping rate $\gamma_{nk}$ cannot be obtained within the framework of the linear Schrödinger equation and is introduced into the linear optical equations phenomenologically [13]; its value (29) is derived only in the framework of quantum electrodynamics. In the theory under consideration, the damping rate (29) is a direct and natural consequence of the nonlinear Schrödinger equation (5).

Let us consider a conventional approach for such systems in the case in which $|\omega_{nk} - \omega| \ll \omega_{nk}$ [13]. Substituting $\cos \omega t = \frac{1}{2}[\exp(i\omega t) + \exp(-i\omega t)]$ and discarding (by averaging) the rapidly oscillating terms, one transforms equations (22), (23) into the form

$$i\frac{dc_n}{dt} = -\frac{1}{2}b_{nk}c_k \exp(i\Omega t) - i\gamma_{nk}c_n|c_k|^2 \tag{31}$$

$$i\frac{dc_k}{dt} = -\frac{1}{2}b_{nk}^*c_n \exp(-i\Omega t) + i\gamma_{nk}c_k|c_n|^2 \tag{32}$$

where

$$\Omega = \omega_{nk} - \omega \tag{33}$$

Accordingly, equations (27) and (28) in this case take the form

$$\frac{d\rho_{nn}}{dt} = -\frac{d\rho_{kk}}{dt} = i\frac{1}{2}[\rho_{kn}b_{nk}\exp(i\Omega t) - \rho_{nk}b_{nk}^*\exp(-i\Omega t)] - 2\gamma_{nk}\rho_{nn}\rho_{kk} \tag{34}$$

$$\frac{d\rho_{nk}}{dt} = \frac{d\rho_{kn}^*}{dt} = (\rho_{kk} - \rho_{nn})[i\frac{1}{2}b_{nk}\exp(i\Omega t) - \gamma_{nk}\rho_{nk}] \tag{35}$$

Let us consider the stationary solution of equations (34) and (35), which corresponds to $\rho_{nn} = const$ and $\rho_{kk} = const$, while the parameters $\rho_{nk}$ and $\rho_{kn}$ will be the oscillating functions. The solution of equations (34) and (35) can be found in the form

$$\rho_{nk} = a \exp(i\Omega t) \tag{36}$$

where $a$ is a constant.

Substituting expression (36) into equations (34) and (35), we obtain the equations

$$i\frac{1}{2}[a^* b_{nk} - a b_{nk}^*] - 2\gamma_{nk}\rho_{nn}\rho_{kk} = 0 \tag{37}$$

$$i\Omega a = (\rho_{kk} - \rho_{nn})(i\frac{1}{2}b_{nk} - \gamma_{nk} a) \tag{38}$$

Hence, using (25), we obtain

$$a = i\frac{1}{2}\frac{b_{nk}(1-2\rho_{nn})}{[\gamma_{nk}(1-2\rho_{nn})+i\Omega]} \tag{39}$$

$$|b_{nk}|^2 \frac{(1-2\rho_{nn})^2}{\gamma_{nk}^2(1-2\rho_{nn})^2+\Omega^2} - 4\rho_{nn}(1-\rho_{nn}) = 0 \tag{40}$$

In particular,

$$|a|^2 = \rho_{nn}(1-\rho_{nn}) \tag{41}$$

Equation (40) can be rewritten in the form

$$(\Omega/\gamma_{nk})^2 = \left(\frac{|b_{nk}|^2/\gamma_{nk}^2}{4\rho_{nn}(1-\rho_{nn})} - 1\right)(1-2\rho_{nn})^2 \tag{42}$$

The solution of equation (42) exists only if its right-hand side is positive. This arrangement is possible only if

$$|b_{nk}|^2/\gamma_{nk}^2 > 4\rho_{nn}(1-\rho_{nn}) \tag{43}$$

One can rewrite this condition in the form

$$\rho_{nn}^2 - \rho_{nn} + \frac{1}{4}|b_{nk}|^2/\gamma_{nk}^2 > 0 \tag{44}$$

which is satisfied for any $\rho_{nn}$ if

$$|b_{nk}|^2/\gamma_{nk}^2 \geq 1 \tag{45}$$

At the same time, if

$$|b_{nk}|^2/\gamma_{nk}^2 < 1 \tag{46}$$

then condition (44) is satisfied only if

$$0 \leq \rho_{nn} \leq \frac{1}{2}\left(1 - \sqrt{1-|b_{nk}|^2/\gamma_{nk}^2}\right) \tag{47}$$

or if

$$\frac{1}{2}\left(1 + \sqrt{1-|b_{nk}|^2/\gamma_{nk}^2}\right) \leq \rho_{nn} \leq 1 \tag{48}$$

The solutions of equation (42) for different values of the parameter $|b_{nk}|/\gamma_{nk}$ are shown in Fig. 1. Equation (42) has two solutions for the same $\Omega$ (see Fig. 1). Their sum is equal to one, which means that one root can be considered to be $\rho_{nn}$, while the other is considered to be $\rho_{kk}$. Theoretically, each of these roots can correspond to $\rho_{nn}$. Recall that condition (14) was accepted. Therefore, due to the spontaneous emission, mode $n$ always loses an electric charge, while mode $k$ receives it. This arrangement means that only the smaller root of equation (42) should correspond to mode $n$, while the larger root corresponds to mode $k$. Such a solution will obviously be always stable: there are no small perturbations that could violate this condition because the system will always return to it. In contrast, the second solution, in which $\rho_{nn}$ corresponds to a larger root of equation (42), will be unstable, and any small perturbations lead to the system spontaneously returning to the first stable state due to crossflow of the electric charge of the electron wave from the upper excited mode to the lower mode. Thus, in the stable state, the smaller of the two roots of equation (42) always corresponds to $\rho_{nn}$, while the larger root corresponds to $\rho_{kk}$.

Then, for the case in (46), the maximum value of $\rho_{nn}$, which can be achieved at resonance ($\Omega = 0$), will be $(\rho_{nn})_{max} = \frac{1}{2}\left(1 - \sqrt{1 - |b_{nk}|^2/\gamma_{nk}^2}\right)$. The corresponding minimum value of $\rho_{kk}$ will be $(\rho_{kk})_{min} = \frac{1}{2}\left(1 + \sqrt{1 - |b_{nk}|^2/\gamma_{nk}^2}\right)$. It should be noted that under condition (46), the solution of equation (42) has a discontinuity at $\Omega = 0$. In fact, at the point $\Omega = 0$ (i.e., at resonance), this equation has a formal solution $\rho_{nn} = \rho_{kk} = 0.5$, but in the limit $\Omega \to 0$, it has a solution $\lim_{\Omega \to 0} \rho_{nn} = \frac{1}{2}\left(1 \pm \sqrt{1 - |b_{nk}|^2/\gamma_{nk}^2}\right)$. The resonant solution (at $\Omega = 0$) can be realized in the experiment only if the condition $\Omega = 0$ is provided exactly. Any deviation from this condition will lead to the "destruction" of the resonant solutions. Because the exact condition $\Omega = 0$ cannot be realized in an experiment due to presence of the fluctuations in any system, it is clear that the resonance solution has no practical interest.

Let us compare the solution of the exact nonlinear equations (34), (35) with the solution of the linear optical equations that are considered in quantum optics [13]. The linear optical equations can be obtained as a linear approximation of equations (34) and (35), which correspond to the case of weak excitation of the upper mode $n$, i.e., when $\rho_{nn} \ll 1$. In our notation, the stationary solution of the linear optical equations with damping due to spontaneous emission has the form [13]

$$\rho_{nn} = \frac{\frac{1}{4}|b_{nk}|^2}{\Omega^2 + \gamma_{nk}^2 + \frac{1}{2}|b_{nk}|^2} \qquad (49)$$

Fig. 2 shows a comparison of the stationary solutions (42) and (49).

In contrast to the nonlinear equations (34) and (35), the linear optical equations have a unique stationary solution $\rho_{nn}$, which corresponds to the smaller (stable) of two stationary solutions of the nonlinear equations (34) and (35).

Fig. 2 shows that the lower (stable) of two stationary solutions of nonlinear optical equations (34) and (35) tends to the stationary solution (49) of linear optical equations only for small values of the parameter $|b_{nk}|/\gamma_{nk} < 0.4$ as well as for the asymptotically large values $|\omega_{nk} - \omega|/\gamma_{nk} \gg 1$. Both correspond to the relatively weak actions of the optical field on the atom. Thus, the linear optical equation with damping due to spontaneous emission can describe only the impact of a weak optical field on the atom when $|b_{nk}|^2/\gamma_{nk}^2 \ll 1$. In the description of strong impacts, especially near the resonance frequency, it is necessary to use the nonlinear optical equations (34) and (35).

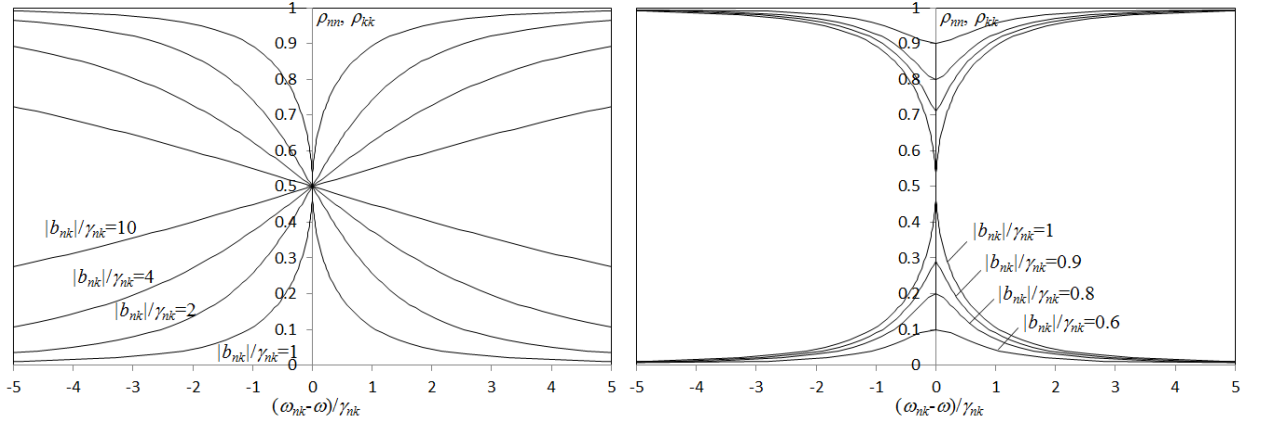

Fig. 1. Dependencies of $\rho_{nn}(\Omega/\gamma_{nk})$ and $\rho_{kk}(\Omega/\gamma_{nk})$ for different values of the parameter $|b_{nk}|/\gamma_{nk}$. The lower branch corresponds to $\rho_{nn}$, and the upper branch corresponds to $\rho_{kk}$.

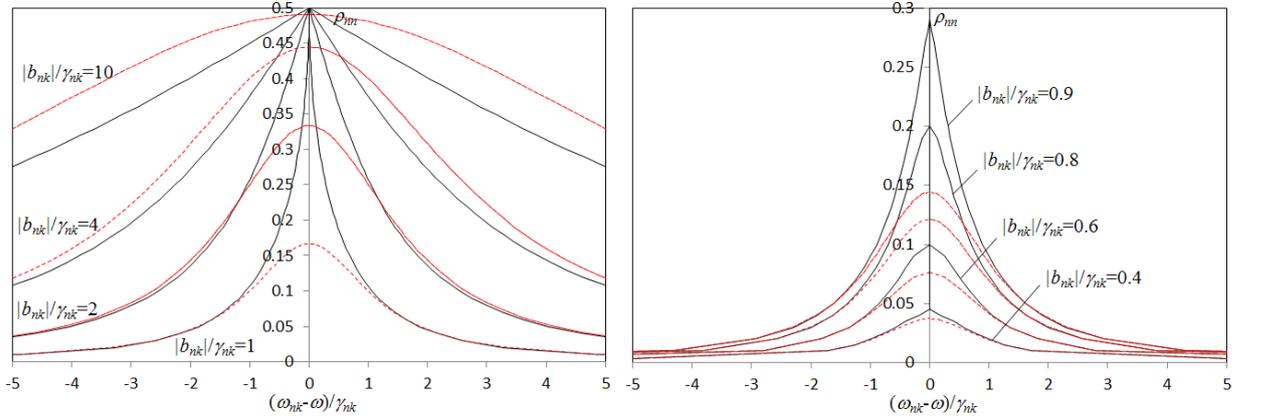

Fig. 2. (Colour online) Comparison of the dependencies $\rho_{nn}(\Omega/\gamma_{nk})$ for different values of parameter $|b_{nk}|/\gamma_{nk}$, which corresponds to stationary solutions of nonlinear optical equations (34) and (35) (solid black lines) and linear optical equations (dashed red lines).

In particular, in a strong light field when $|b_{nk}|^2/\gamma_{nk}^2 \geq 1$, at resonance ($\Omega = 0$), a simultaneous strong excitation of both modes at which $\rho_{nn}(0) = \rho_{kk}(0) = \frac{1}{2}$ occurs. In this case, a saturation occurs when the value $\rho_{nn}(0) = \rho_{kk}(0)$ does not depend on the intensity of exposure to the atom $|b_{nk}|^2/\gamma_{nk}^2$. At the same time, at the weak excitation when $|b_{nk}|^2/\gamma_{nk}^2 < 1$, it will always be $\rho_{nn} < \rho_{kk}$, even under resonance conditions.

Let us estimate the ratio $\gamma_{nk}/\omega_{nk} = \frac{2\omega_{nk}^2}{3\hbar c^3}|\mathbf{d}_{nk}|^2$, while accounting for $|\mathbf{d}_{nk}|\sim ea_B$ and $a_B = \frac{\hbar^2}{me^2}$, $\omega_{nk} \sim \frac{me^4}{\hbar^3}$. As a result, we obtain

$$\gamma_{nk}/\omega_{nk} \sim \alpha^3 \ll 1$$

where $\alpha = \frac{e^2}{\hbar c}$ is the fine-structure constant.

It follows from Figs. 1 and 2 that when $|b_{nk}|/\gamma_{nk} \sim 1$, and even more so when $|b_{nk}|/\gamma_{nk} < 1$, the width of the frequency range $\Delta\Omega$ on which a noticeable change in the parameters $c_n$ occurs is of the order of $\Delta\Omega \sim \gamma_{nk}$, and therefore, $\Delta\Omega \ll \omega_{nk}$. Using expressions (36) and (39), we obtain $|\dot{c}_n| \sim |c_n|\Omega|b_{nk}|/\gamma_{nk}$. As a result, we can conclude that in this case, $|\dot{c}_n| \ll \omega_{nk}|c_n|$. At the same time, when $|b_{nk}|/\gamma_{nk} \gg 1$, the width of the frequency range $\Delta\Omega \gg \gamma_{nk}$, and in this case, it could be $|\dot{c}_n| \sim \omega_{nk}|c_n|$ or even $|\dot{c}_n| > \omega_{nk}|c_n|$. Thus, we conclude that condition (18) can be violated only for very strong electromagnetic waves, for which $|b_{nk}|/\gamma_{nk} \gg 1$. Such an electromagnetic wave can cause a so-called tunnel ionization of the atom, which is not considered here.

## 3 Light scattering by an atom

Using the results of the previous section, it is easy to calculate the secondary radiation (induced and spontaneous), which an atom that is in the field of a classical electromagnetic wave creates. This radiation will be perceived as the scattering of the incident electromagnetic wave. The intensity of the electric dipole radiation according to classical electrodynamics is defined by the expression [12]

$$I = \frac{2}{3c^3}\overline{\ddot{\mathbf{d}}^2} \qquad (50)$$

where $\mathbf{d}$ is the electric dipole moment of the electron wave in a hydrogen atom; the bar denotes averaging over time.

For a two-level atom, as discussed in the previous section, accounting for expressions (17), (18) and (24) implies that

$$\ddot{\mathbf{d}} = -\omega_{nk}^2[\rho_{kn}\mathbf{d}_{nk}\exp(i\omega_{nk}t) + \rho_{nk}\mathbf{d}_{nk}^*\exp(-i\omega_{nk}t)] \qquad (51)$$

If the atom is in a stationary forced excited state, then the parameter $\rho_{nk}$ is determined by the expressions (36) and (39), and it oscillates at the frequency given in (33).

In this case,

$$\ddot{\mathbf{d}} = -\omega_{nk}^2[a^*\mathbf{d}_{nk}\exp(i\omega t) + a\mathbf{d}_{nk}^*\exp(-i\omega t)] \qquad (52)$$

and for the intensity of the scattered radiation, we obtain

$$I = \frac{4\omega_{nk}^4}{3c^3}|a|^2|\mathbf{d}_{nk}|^2 \tag{53}$$

or when accounting for expressions (29) and (41), we obtain

$$I = 2\gamma_{nk}\rho_{nn}(1-\rho_{nn})\hbar\omega_{nk} \tag{54}$$

According to the expression in (52), in the approximation under consideration, we are contending with Rayleigh scattering. However, if we account for the rapidly oscillating terms in equations (27) and (28), the non-Rayleigh components in the scattering spectrum will be detected, but their intensity will be negligible.

Let us calculate the scattering cross-section

$$d\sigma = \frac{dI}{\overline{S}} \tag{55}$$

where [12]

$$dI = \frac{\overline{\ddot{\mathbf{d}}^2}}{4\pi c^3}\sin^2\theta\, do \tag{56}$$

is the amount of energy that is emitted (scattered) by the atom per unit time per unit solid angle $do$; $\theta$ is the angle between the vector $\ddot{\mathbf{d}}$ and the direction of the scattering; and

$$\overline{S} = \frac{c}{4\pi}\overline{|\mathbf{E}|^2} \tag{57}$$

is the energy flux density of the incident electromagnetic wave.

In our case,

$$\overline{S} = \frac{c}{8\pi}|\mathbf{E}_0|^2 \tag{58}$$

Then, accounting for expressions (29) and (41), we obtain

$$d\sigma = 6\gamma_{nk}\frac{\hbar\omega_{nk}}{c|\mathbf{E}_0|^2}\rho_{nn}(1-\rho_{nn})\sin^2\theta\, do \tag{59}$$

Using expressions (29) and (40), we can also write

$$d\sigma = \frac{9\hbar^2 c^2\gamma_{nk}^2}{4\omega_{nk}^2}\frac{|b_{nk}|^2}{|\mathbf{d}_{nk}|^2|\mathbf{E}_0|^2}\frac{(1-2\rho_{nn})^2}{[\gamma_{nk}^2(1-2\rho_{nn})^2+\Omega^2]}\sin^2\theta\, do \tag{60}$$

The scattering pattern will be determined by the parameter $b_{nk}$.

In that case, when the vector $\mathbf{d}_{nk}$ is real-valued, based on definition (30), we can write

$$|b_{nk}|^2 = \frac{1}{\hbar^2}|\mathbf{d}_{nk}|^2|\mathbf{E}_0|^2\cos^2\vartheta \tag{61}$$

where $\vartheta$ is the angle between the vectors $\mathbf{d}_{nk}$ and $\mathbf{E}_0$.

Then, we obtain

$$d\sigma = \frac{9c^2\gamma_{nk}^2}{4\omega_{nk}^2}\frac{(1-2\rho_{nn})^2}{[\gamma_{nk}^2(1-2\rho_{nn})^2+\Omega^2]}\sin^2\theta\cos^2\vartheta\, do \tag{62}$$

In particular, at the resonance frequency ($\Omega = 0$),

$$d\sigma = \frac{9c^2}{4\omega_{nk}^2}\sin^2\theta\cos^2\vartheta\, do \tag{63}$$

In other cases ($\Omega \neq 0$), in the calculation of the right-hand side of expression (62), it is necessary to account for the fact that parameter $\rho_{nn}$, which is a solution of equation (42), will depend on the nondimensional parameter $|b_{nk}|^2/\gamma_{nk}^2$ (see Fig. 1). Accounting for expressions (61) and (29), we obtain

$$|b_{nk}|^2/\gamma_{nk}^2 = \frac{9c^6|\mathbf{E}_0|^2}{4\omega_{nk}^6|\mathbf{d}_{nk}|^2} \cos^2 \vartheta \tag{64}$$

At a low intensity of the incident electromagnetic wave, when

$$\frac{9c^6|\mathbf{E}_0|^2}{4\omega_{nk}^6|\mathbf{d}_{nk}|^2} \ll 1 \tag{65}$$

we obtain $\rho_{nn} \ll 1$, and then,

$$d\sigma \approx \frac{9c^2}{4\omega_{nk}^2} \frac{\gamma_{nk}^2}{(\gamma_{nk}^2+\Omega^2)} \sin^2 \theta \cos^2 \vartheta \, do \tag{66}$$

In cases when the vector $\mathbf{d}_{nk}$ is complex-valued, we can write $\mathbf{d}_{nk} = \mathbf{d}_{nk}^{(r)} + i\mathbf{d}_{nk}^{(i)}$, where $\mathbf{d}_{nk}^{(r)}$ and $\mathbf{d}_{nk}^{(i)}$ are real-valued vectors. In this case, it is possible to introduce the angle $\chi$, such that $\left|\mathbf{d}_{nk}^{(r)}\right|/|\mathbf{d}_{nk}| = \cos\chi$ and $\left|\mathbf{d}_{nk}^{(i)}\right|/|\mathbf{d}_{nk}| = \sin\chi$. Then, the relation (30) can be written as $b_{nk} = b_{kn}^* = \frac{1}{\hbar}|\mathbf{d}_{nk}||\mathbf{E}_0|(\cos\chi \cos\vartheta_r + i\sin\chi \cos\vartheta_i)$, where $\vartheta_r$ and $\vartheta_i$ are the angles between the vector $\mathbf{E}_0$ and the vectors $\mathbf{d}_{nk}^{(r)}$ and $\mathbf{d}_{nk}^{(i)}$, respectively. Hence, we obtain

$$|b_{nk}|^2 = \frac{1}{\hbar^2}|\mathbf{d}_{nk}|^2|\mathbf{E}_0|^2(\cos^2\chi \cos^2\vartheta_r + \sin^2\chi \cos^2\vartheta_i) \tag{67}$$

Then, using expression (60), we obtain

$$d\sigma = \frac{9c^2\gamma_{nk}^2}{4\omega_{nk}^2} \frac{(1-2\rho_{nn})^2}{[\gamma_{nk}^2(1-2\rho_{nn})^2+\Omega^2]} \sin^2\theta \, (\cos^2\chi \cos^2\vartheta_r + \sin^2\chi \cos^2\vartheta_i) do \tag{68}$$

In particular, at the resonance frequency ($\Omega = 0$),

$$d\sigma = \frac{9c^2}{4\omega_{nk}^2} \sin^2\theta \, (\cos^2\chi \cos^2\vartheta_r + \sin^2\chi \cos^2\vartheta_i) do \tag{69}$$

In other cases ($\Omega \neq 0$), in the calculation of the right-hand side of expression (68), it is necessary to account for the fact that the parameter $\rho_{nn}$, which is a solution of equation (42), will depend on the nondimensional parameter $|b_{nk}|^2/\gamma_{nk}^2$ (see Fig. 1). Using the expressions (67) and (29), we obtain in this case

$$|b_{nk}|^2/\gamma_{nk}^2 = \frac{9c^6|\mathbf{E}_0|^2}{4\omega_{nk}^6|\mathbf{d}_{nk}|^2}(\cos^2\chi \cos^2\vartheta_r + \sin^2\chi \cos^2\vartheta_i) \tag{70}$$

At a low intensity of the incident electromagnetic wave, when condition (65) is satisfied and $\rho_{nn} \ll 1$, we obtain

$$d\sigma \approx \frac{9c^2}{4\omega_{nk}^2} \frac{\gamma_{nk}^2}{[\gamma_{nk}^2+\Omega^2]} \sin^2\theta \, (\cos^2\chi \cos^2\vartheta_r + \sin^2\chi \cos^2\vartheta_i) do \tag{71}$$

Thus, the Rayleigh scattering of an electromagnetic wave by a hydrogen atom is fully described within the framework of classical field theory without any quantization.

# 4 Photoelectric effect

Until now, we have assumed that under the influence of an incident electromagnetic wave, the electron wave in an atom is only redistributed between its eigenmodes but not emitted outward by the atom. In this case, internal electric currents arise inside the atom that, however, cannot be detected by macroscopic devices. Such a situation occurs at a relatively low frequency of the incident electromagnetic wave. If this frequency is sufficiently large, then an emission of the electron wave by the atom occurs. Because the electron wave has an electric charge that is continuously distributed in space [10,11], in this case, an external electric current (photoelectric current) appears that can be detected by macroscopic devices. As a result, the photoelectric effect will be observed.

The photoelectric effect has a special place in quantum mythology because it became the first physical effect, for explanation of which the quantization of light was introduced (A. Einstein, 1905).

By the early 20th century, the three basic laws of the photoelectric effect were experimentally established: (i) the photoelectric current is proportional to the intensity of incident light; (ii) the maximum kinetic energy of the emitted photoelectrons varies linearly with the frequency of incident electromagnetic radiation and does not depend on the flux; and (iii) for each substance, there is a threshold frequency (the so-called red edge of the photoelectric effect), below which the photoelectric current is not observed.

The second and third laws of the photoelectric effect would appear to contradict classical electrodynamics, which requires dependence of the kinetic energy of the emitted photoelectrons on the intensity of the incident light. Such a conclusion necessarily follows from the analysis of the motion of charged particles - electrons in the field of a classical electromagnetic wave. Thus, the attempts to explain the photoelectric effect within the framework of classical mechanics and classical electrodynamics were unsuccessful.

This contradiction was overcome due to the quantization of radiation, which postulates that the absorption of light occurs in the form of discrete quanta $\hbar\omega$ (A. Einstein, 1905). At present, in connection with this finding, it is considered to be generally accepted that the photoelectric effect provides "evidence" for the quantum nature of light.

However, in the early years of quantum mechanics, it was shown that the photoelectric effect is fully described within the framework of so-called semiclassical theory, in which light is considered to be a classical electromagnetic wave, while the atom is quantized and described by the wave equation, e.g., the Schrödinger equation or the Dirac equation [1,13,15,16]. In this case, the wave equation is solved as a typical classical field equation, whereby a continuous wave field is calculated. A "quantization" of this wave field occurs only at the stage of interpreting the solution, from which the "probability of photoelectron emission" from an atom is determined. Because the electron in such a theory is considered to be a quantum particle and light is considered to be a classical electromagnetic field, such a theory is considered to be "semi-classical".

However, as shown in [7-11], there is no need to introduce the quantization of electromagnetic and electron fields because this interpretation is external to the wave equation, and it does not follow from these equations. Moreover, this approach is superfluous in explaining the many physical phenomena that before were interpreted as a result of the quantization of matter.

As will be shown below, the failures of classical electrodynamics in explaining the photoelectric effect are connected with the incorrect postulate that electrons are particles. I will show that for a consistent explanation of the photoelectric effect within the framework of classical field theory, it is sufficient to abandon this postulate and instead consider continuous classical electron waves instead of the particles-electrons [10,11]. The considered theory is fully classical because it does not contain not only the quantization of the radiation but also the quantization of the electron wave.

From the considered point of view [7-11], the photoelectric effect represents an emission of the continuous charged electron wave by an atom that was excited by the incident classical electromagnetic wave. Formally, the photoelectric effect is no different from the stimulated emission of electromagnetic waves by an atom [11], with the only difference being that the electron wave emitted by an atom is electrically charged while the electromagnetic wave does not carry the electric charge. Assuming that the electric charge is continuously distributed in the electron wave [10,11], one concludes that in the process of the emission of the electron wave, the atom is positively charged continuously. However, accounting for the fact that the electron wave for an as yet inexplicable reason does not "feel" its own electrostatic field [11], this process will not affect the emission of the following "portions" of the continuous electron wave because they must overcome the same electrostatic potential of the nucleus.

Let us consider the photoelectric effect for the hydrogen atom being in the classic monochromatic electromagnetic wave.

In this section, we neglect the inverse action on the electron wave of its own non-stationary electromagnetic field. For this reason, the last term in the Schrödinger equation (5), which is associated with a spontaneous emission of the electromagnetic waves, will not be considered, and we will use the conventional linear Schrödinger equation

$$i\hbar \frac{\partial \psi}{\partial t} = -\frac{1}{2m_e}\Delta\psi - \frac{e^2}{r}\psi + \psi e \mathbf{r} \mathbf{E}_0 \cos \omega_0 t \tag{72}$$

where $\omega_0$ is the frequency of the incident light. We will consider here the approximation in (3), when the wavelength of the incident electromagnetic wave is substantially larger than the characteristic spatial size of the electron field in the hydrogen atom, which is of the order of the Bohr radius $a_B$.

The wave function of an electron wave can be represented as in [17]

$$\psi = \sum_k c_k(t)u_k(\mathbf{r})\exp(-i\omega_k t) + \sum_n \int_0^\infty C_n(\omega,t)f_n(\mathbf{r},\omega)\exp(-i\omega t)\,d\omega \qquad (73)$$

where the first sum describes that part of the electron wave that is contained in the eigenmodes of the atom (i.e., corresponding to a "finite motion" of the electron wave), and for this term, all $\omega_k < 0$, while the integrals describe the electron waves that are emitted by an atom (i.e., which corresponds to the "infinite motion" of the electron wave), to which it is known that $\omega > 0$ corresponds. The indices $n$ and $k$ run through the appropriate integer values. The functions $u_k(\mathbf{r})$ and $f_n(\mathbf{r},\omega)$ are the eigenfunctions of the stationary Schrödinger equation (7), while the frequencies $\omega_k$ are the eigenvalues that correspond to the eigenfunctions $u_k(\mathbf{r})$.

The eigenfunctions $u_k(\mathbf{r})$ and $f_n(\mathbf{r},\omega)$ satisfy the orthogonality conditions

$$\int u_k(\mathbf{r})u_n^*(\mathbf{r})dV = \delta_{nk} \qquad (74)$$

$$\int f_k(\mathbf{r},\omega')f_n^*(\mathbf{r},\omega'')dV = \delta_{nk}\delta(\omega'-\omega'') \qquad (75)$$

$$\int u_k(\mathbf{r})f_n^*(\mathbf{r},\omega'')dV = 0 \qquad (76)$$

Substituting expression (73) into equation (72) and using the orthogonality conditions (74)-(76), we obtain

$$i\hbar\dot{c}_k(t)\exp(-i\omega_k t) =$$
$$e\mathbf{E}_0\cos\omega_0 t \sum_n c_n(t)\int \mathbf{r}u_n(\mathbf{r})u_k^*dV \exp(-i\omega_n t) +$$
$$e\mathbf{E}_0\cos\omega_0 t \sum_n \int_0^\infty C_n(\omega,t)\int \mathbf{r}u_k^* f_n(\mathbf{r},\omega)dV \exp(-i\omega t)\,d\omega \qquad (77)$$

and

$$i\hbar\dot{C}_n(\omega,t)\exp(-i\omega t) = e\mathbf{E}_0\cos\omega_0 t \sum_k c_k(t)\int \mathbf{r}f_n^*(\mathbf{r},\omega)u_k(\mathbf{r})dV \exp(-i\omega_k t) +$$
$$e\mathbf{E}_0\cos\omega_0 t \sum_k \int_0^\infty C_k(\omega',t)\int \mathbf{r}f_k(\mathbf{r},\omega')f_n^*(\mathbf{r},\omega)dV \exp(-i\omega' t)\,d\omega' \qquad (78)$$

Within the framework of perturbation theory and assuming that all of the modes of the electron wave (both discrete and continuous), except for the ground mode $u_1$, are weakly excited, we obtain

$$i\hbar\dot{c}_1(t)\exp(-i\omega_1 t) =$$
$$-c_1(t)(\mathbf{E}_0\mathbf{d}_{11})\cos\omega_0 t \exp(-i\omega_1 t) +$$
$$e\mathbf{E}_0\cos\omega_0 t \sum_n \int_0^\infty C_n(\omega,t)\int \mathbf{r}u_1^* f_n(\mathbf{r},\omega)dV \exp(-i\omega t)\,d\omega \qquad (79)$$

$$i\hbar \dot{C}_n(\omega, t) =$$
$$e\mathbf{E}_0 \cos \omega_0 t\, c_1(t) \exp[-i(\omega_1 - \omega)t] \int \mathbf{r} f_n^*(\mathbf{r}, \omega) u_1(\mathbf{r}) dV +$$
$$e\mathbf{E}_0 \cos \omega_0 t \exp(i\omega t) \sum_k \int_0^\infty C_k(\omega', t) \int \mathbf{r} f_k(\mathbf{r}, \omega') f_n^*(\mathbf{r}, \omega) dV \exp(-i\omega' t)\, d\omega' \quad (80)$$

where
$$\mathbf{d}_{nk} = -e \int \mathbf{r} u_n(\mathbf{r}) u_k^* dV \quad (81)$$

For a weak electromagnetic wave, which causes weak excitation of an atom, $c_1 \approx 1$. For this reason, we can discard terms in equation (80) that contain $\mathbf{E}_0 C_k(\omega', t)$, as small of the second order. In equation (79), these terms cannot be discarded because the change in $c_1$ will have a second order in $\mathbf{E}_0$. Then, we obtain

$$i\hbar \dot{c}_1(t) = -(\mathbf{E}_0 \mathbf{d}_{11}) \cos \omega_0 t + \tfrac{1}{2} e\mathbf{E}_0 \sum_n \int_0^\infty C_n(\omega, t) \mathbf{U}_{1n}(\omega) \exp[-i(\omega - \omega_1 - \omega_0)t]\, d\omega +$$
$$\tfrac{1}{2} e\mathbf{E}_0 \sum_n \int_0^\infty C_n(\omega, t) \mathbf{U}_{1n}(\omega) \exp[-i(\omega - \omega_1 + \omega_0)t]\, d\omega \quad (82)$$

$$i\hbar \dot{C}_n(\omega, t) = \tfrac{1}{2} e\mathbf{E}_0 \cdot \mathbf{U}_{1n}^*(\omega) \exp[-i(\omega_1 - \omega - \omega_0)t] + \tfrac{1}{2} e\mathbf{E}_0 \cdot \mathbf{U}_{1n}^*(\omega) \exp[-i(\omega_1 - \omega + \omega_0)t] \quad (83)$$

where
$$\mathbf{U}_{1n}(\omega) = \int \mathbf{r} u_1^*(\mathbf{r}) f_n(\mathbf{r}, \omega) dV \quad (84)$$

Neglecting the purely oscillatory term $-(\mathbf{E}_0 \mathbf{d}_{11}) \cos \omega_0 t$ in equation (82) (which can be accomplished, for example, by averaging equation (82) over rapid oscillations with a frequency $\omega_0$), we obtain

$$i\hbar \dot{c}_1(t) =$$
$$\tfrac{1}{2} e\mathbf{E}_0 \sum_n \int_0^\infty C_n(\omega, t) \mathbf{U}_{1n}(\omega) \exp[-i(\omega - \omega_1 - \omega_0)t]\, d\omega +$$
$$\tfrac{1}{2} e\mathbf{E}_0 \sum_n \int_0^\infty C_n(\omega, t) \mathbf{U}_{1n}(\omega) \exp[-i(\omega - \omega_1 + \omega_0)t]\, d\omega \quad (85)$$

Integrating equation (83) with respect to the time from zero to f, we obtain

$$C_n(\omega, t) = \frac{e}{2\hbar} \frac{\exp[-i(\omega_1 - \omega - \omega_0)t] - 1}{(\omega_1 - \omega - \omega_0)} \mathbf{E}_0 \cdot \mathbf{U}_{1n}^*(\omega) + \frac{e}{2\hbar} \frac{\exp[-i(\omega_1 - \omega + \omega_0)t] - 1}{(\omega_1 - \omega + \omega_0)} \mathbf{E}_0 \cdot \mathbf{U}_{1n}^*(\omega) \quad (86)$$

Because the frequencies have $\omega_0 > 0$, $\omega > 0$ and $\omega_1 < 0$, the value $\omega_1 - \omega - \omega_0$ is not equal to zero for any $\omega$, and thus, the first term will always be limited and will describe the oscillations that are of small amplitude. At the same time, $\omega_1 - \omega + \omega_0 = 0$ at the resonance frequency of $\omega_0 = |\omega_1| + \omega$, and near the resonant frequency, the second term in (86) will increase indefinitely. Therefore, the second term in (86) makes the main contribution to the effect that is under consideration. Neglecting the first term in expression (86), we obtain

$$C_n(\omega, t) = \frac{e}{2\hbar} \frac{\exp[-i(\omega_1 - \omega + \omega_0)t] - 1}{(\omega_1 - \omega + \omega_0)} \mathbf{E}_0 \cdot \mathbf{U}_{1n}^*(\omega) \quad (87)$$

Let us calculate the photoelectric current that arises upon excitation of the atom by the incident electromagnetic wave.

This goal can be accomplished by calculating the electric current density according to the formula

$$\mathbf{j} = i\frac{ec^2}{2\omega_e}(\psi^*\nabla\psi - \psi\nabla\psi^*) - \frac{e^2c}{\hbar\omega_e}\mathbf{A}\psi\psi^* \qquad (88)$$

and integrating it over the surface of an infinite sphere whose centre is in the nucleus of the atom. However, it is more convenient to accomplish this step while using the law of conservation of charge and accounting for the fact that $q_k = -e|c_k|^2$ is the electric charge that is contained in mode $k$ of the electron wave [11]. Then, $\dot{q}_k$ is the internal electric current in the atom, by which mode $k$ is exchanged with all of the other modes of the electron wave (including continuous modes, if they exist), i.e., the amount of electric charge of the electron wave, which goes into mode $k$ from other modes or goes out of mode $k$ into other modes, per unit time. Because in this case it is considered that only one (ground) eigenmode $u_1$ of the hydrogen atom is excited, then the photoelectric current

$$I_{ph} = -\dot{q}_1 \qquad (89)$$

or

$$I_{ph} = -e\frac{d|c_1|^2}{dt} \qquad (90)$$

Using equation (85), we obtain the same approximation

$$I_{ph} = -\frac{e^2}{2i\hbar}\sum_n \int_0^\infty C_n(\omega,t)\mathbf{E}_0 \cdot \mathbf{U}_{1n}(\omega)\exp[-i(\omega-\omega_1-\omega_0)t]\,d\omega - \frac{e^2}{2i\hbar}\sum_n \int_0^\infty C_n(\omega,t)\mathbf{E}_0 \cdot \mathbf{U}_{1n}(\omega)\exp[-i(\omega-\omega_1+\omega_0)t]\,d\omega + \frac{e^2}{2i\hbar}\sum_n \int_0^\infty C_n^*(\omega,t)\mathbf{E}_0 \cdot \mathbf{U}_{1n}^*(\omega)\exp[i(\omega-\omega_1-\omega_0)t]\,d\omega + \frac{e^2}{2i\hbar}\sum_n \int_0^\infty C_n^*(\omega,t)\mathbf{E}_0 \cdot \mathbf{U}_{1n}^*(\omega)\exp[i(\omega-\omega_1+\omega_0)t]\,d\omega \qquad (91)$$

Substituting $C_n(\omega,t)$ from (87) into expression (91), we obtain

$$I_{ph} = \frac{e^3}{2\hbar^3}\sum_n \int_0^\infty \frac{\sin[(\omega-\omega_1-\omega_0)t]}{(\omega-\omega_1-\omega_0)}(\mathbf{E}_0 \cdot \mathbf{U}_{1n}^*)(\mathbf{E}_0 \cdot \mathbf{U}_{1n})\,d\omega + \frac{e^3}{4i\hbar^3}\exp(-2i\omega_0 t)\sum_n \int_0^\infty \frac{1-\exp[-i(\omega-\omega_1-\omega_0)t]}{(\omega-\omega_1-\omega_0)}(\mathbf{E}_0 \cdot \mathbf{U}_{1n}^*)(\mathbf{E}_0 \cdot \mathbf{U}_{1n})\,d\omega - \frac{e^3}{4i\hbar^3}\exp(2i\omega_0 t)\sum_n \int_0^\infty \frac{1-\exp[i(\omega-\omega_1-\omega_0)t]}{(\omega-\omega_1-\omega_0)}(\mathbf{E}_0 \cdot \mathbf{U}_{1n})(\mathbf{E}_0 \cdot \mathbf{U}_{1n}^*)\,d\omega \qquad (92)$$

The second and third terms on the right-hand side of expression (92) are rapidly oscillating at a frequency of $\omega_0$, and they can be discarded by averaging over the fast oscillations. Then, we obtain

$$I_{ph} = \frac{e^3}{2\hbar^2}\int_0^\infty \frac{\sin[(\omega-\omega_1-\omega_0)t]}{(\omega-\omega_1-\omega_0)}\sum_n(\mathbf{E}_0 \cdot \mathbf{U}_{1n}^*)(\mathbf{E}_0 \cdot \mathbf{U}_{1n})\,d\omega \qquad (93)$$

Assuming that all of the orientations of the atom in space are equally probable, and therefore the vector $\mathbf{U}_{1n}$ is statistically isotropic, one averages the current (93) over all possible orientations of the atom.

Then,

$$\overline{(\mathbf{E}_0\mathbf{U}_{1n})(\mathbf{E}_0\mathbf{U}_{1n}^*)} = E_{0i}E_{0j}\overline{U_{1n,i}U_{1n,j}^*} \qquad (94)$$

where the bar denotes averaging over all possible orientations, and the indices $i$ and $j$ are the vector indexes.

For the isotropic vector $\mathbf{U}_{1n}$,

$$\overline{U_{1n,i}U_{1n,j}^*} = \tfrac{1}{3}|\mathbf{U}_{1n}|^2 \delta_{ij} \qquad (95)$$

Then,

$$\overline{(\mathbf{E}_0\mathbf{U}_{1n})(\mathbf{E}_0\mathbf{U}_{1n}^*)} = \tfrac{1}{3}|\mathbf{E}_0|^2|\mathbf{U}_{1n}|^2 \qquad (96)$$

Accordingly, for the mean photoelectric current (93), we obtain

$$\overline{I_{ph}} = \beta |\mathbf{E}_0|^2 \qquad (97)$$

where the parameter

$$\beta = \frac{e^3}{6\hbar^2} \int_0^\infty \frac{\sin[(\omega-\omega_1-\omega_0)t]}{(\omega-\omega_1-\omega_0)} \sum_n |\mathbf{U}_{1n}(\omega)|^2 \, d\omega \qquad (98)$$

does not depend on the incident light intensity $|\mathbf{E}_0|^2$, and instead, the parameter $\beta$ depends on the frequency $\omega_0$ of the incident light.

Thus, we have obtained the first law of the photoelectric effect without using the photon hypothesis within the framework of only classical field theory while considering the electromagnetic and electron waves as classical fields.

Let us consider the dependence of the parameter $\beta$ on the frequency of the incident light $\omega_0$.

Let us denote

$$F(\omega) = \sum_n |\mathbf{U}_{1n}(\omega)|^2 \qquad (99)$$

$$x = \omega - \omega_1 - \omega_0 \qquad (100)$$

Then, we obtain

$$\beta = \frac{e^3}{6\hbar^2} \int_{|\omega_1|-\omega_0}^\infty \frac{\sin(xt)}{x} F(x - |\omega_1| + \omega_0) dx \qquad (101)$$

Here, we account for the fact that $\omega_1 < 0$.

The function $\frac{\sin(xt)}{x}$ has a sharp peak in the vicinity of $x = 0$ and has a width of $\Delta x \sim \pi/t$, and at $t \to \infty$, it behaves similar to a delta-function: $\int_{-\infty}^\infty \frac{\sin(xt)}{x} dx = \pi$. The function $F(\omega)$ in the vicinity of $x = 0$ is smooth and varies weakly on the interval $\Delta x \sim \pi/t$.

Therefore, with reasonable accuracy at $\omega_0 < |\omega_1| - \frac{\pi}{2t}$, we can write

$$\beta \approx \frac{e^3}{6\hbar^2} F(0) \int_{|\omega_1|-\omega_0}^{\infty} \frac{\sin(xt)}{x} dx \tag{102}$$

At the same time, at $\omega_0 - |\omega_1| \gg \frac{\pi}{2t}$, it is necessary to account for the fact that a small neighbourhood of the point $x = 0$ will make the main contribution to the integral in (101) (due to the delta-like behaviour of the integrand). As a result, for $\omega_0 - |\omega_1| \gg \frac{\pi}{2t}$, we obtain

$$\beta \approx \frac{\pi e^3}{6\hbar^2} F(\omega_0 - |\omega_1|) \tag{103}$$

In this case, the parameter $\beta$ will vary with the frequency of the incident light $\omega_0$.

Fig. 3 shows, in a nondimensional form, the dependence of the parameter $\beta$ on the frequency difference $\omega_0 - |\omega_1|$ in the vicinity of the frequency $\omega_0 = |\omega_1|$.

We can see that the parameter $\beta$ is virtually zero at $\omega_0 < |\omega_1| - \frac{\pi}{2t}$, and it almost linearly varies from zero to $\frac{\pi e^3}{6\hbar^2} F(0)$ when $\omega_0$ changes in the range from $|\omega_1| - \frac{\pi}{2t}$ to $|\omega_1| + \frac{\pi}{2t}$, and it virtually equals the value in (103) at $\omega_0 > |\omega_1| + \frac{\pi}{2t}$. The width of the frequency range in which there is a noticeable change in the parameter $\beta$ is $\Delta\omega_0 \sim \pi/t$.

Assuming $|\omega_1| \sim 10^{14}$ rad/s (which corresponds to visible light) for the observation time $t > 10^{-9}$s, we obtain $\Delta\omega_0 < 3 \cdot 10^9$ rad/s, which is significantly less than $|\omega_1|$:

$$\Delta\omega_0 \ll |\omega_1| \tag{104}$$

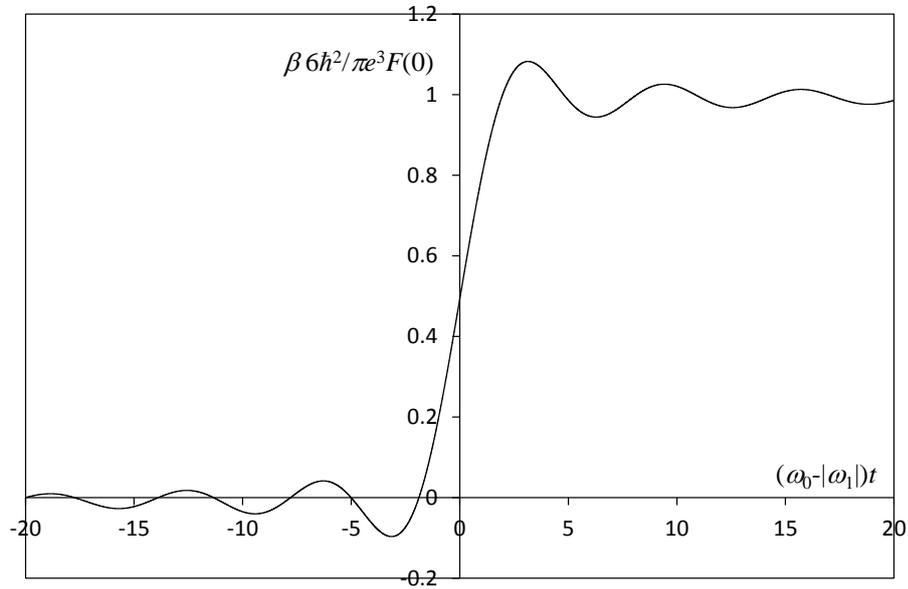

Fig. 3. The dependence of the parameter $\beta$ on the frequency difference $\omega_0 - |\omega_1|$ in the vicinity of the frequency $\omega_0 = |\omega_1|$.

From this analysis, it follows that for the actual duration of the observation, the parameter $\beta$ will have almost a threshold dependence on the frequency of the incident light $\omega_0$: for $\omega_0 < |\omega_1|$, we obtain $\beta \approx 0$, and the photoelectric current is almost absent, while at $\omega_0 > |\omega_1|$, the parameter $\beta$ will take the value in (103), and the photoelectric current (97) will be proportional to the intensity of the incident light.

Thus, we have obtained the third law of the photoelectric effect also without using the photon hypothesis, within only the framework of classical field theory.

Let us now consider the second law of the photoelectric effect. In its conventional form, it establishes the dependence of the kinetic energy of the emitted photoelectrons on the frequency and intensity of the incident radiation. However, in the experiments on the photoelectric effect, the kinetic energy of the photoelectron is not measured directly; it is determined indirectly through the measured stopping potential. Therefore, such wording of the second law of the photoelectric effect already contains some interpretation of the experimental facts; in particular, it assumes that the electrons are indivisible particles that, at the time of escape from the atom, have a definite kinetic energy. In this case, the kinetic energy of the photoelectrons can be determined through the stopping potential at which the photoelectric current is terminated.

Because in the papers of this series we doubt that electrons are particles, it does not make sense to talk about the kinetic energy of the electrons, and we will need a different formulation of the second law of the photoelectric effect.

To rule out any interpretation of the experimental data, the wording of the second law of the photoelectric effect (and in general, of any laws) should use only measured parameters. From this perspective, an *objective formulation of the second law of the photoelectric effect* will be as follows: the stopping potential varies linearly with the frequency of the incident electromagnetic radiation and does not depend on the flux.

Let us consider the function in (87). The square of its modulus $|C_n(\omega, t)|^2$ determines the density of the photoelectric current (88). This function reaches its maximum when

$$\omega_1 - \omega + \omega_0 = 0 \qquad (105)$$

and for large $t$, the largest part of the photoelectric current falls on the narrow range of the frequencies of the electron wave that have the width

$$\Delta\omega \sim \pi/t \qquad (106)$$

near the frequency

$$\omega = \omega_0 - |\omega_1| \qquad (107)$$

When accounting for the smallness of the frequency range (106), it can be assumed that the electron wave that is emitted by an atom is almost monochromatic and has the frequency in (107), which linearly depends on the frequency of the incident light $\omega_0$ and does not depend on its intensity.

Let us place on the path of the electron wave a decelerating potential. In this case, we come to the problem of propagation of the electron wave in the field of the decelerating potential, which is quite accurately described by the linear Schrödinger equation. At large distances from the atom, the electron wave can be considered to be approximately flat. To simplify the analysis, instead of the decelerating potential, having a linear dependence on the coordinates along which the electron wave propagates, let us consider the potential step (barrier) of the same "height" $U_0$ and the same width $L$ to be the actual decelerating potential. The solution of the Schrödinger equation for the potential step is well known [17]: at $\hbar\omega > U_0$, the electron wave passes through a potential step and is partially reflected from it, while when $\hbar\omega < U_0$, the electron wave is mainly reflected from the potential step, although a small part goes through the potential step due to tunnelling. The transmission coefficient of the electron wave for the potential step (in our interpretation, this coefficient is the ratio of the electric current of the electron wave behind the potential step to the electric current of the electron wave arriving to the potential steps from an atom) in the limiting case $\hbar\omega = U_0$ is defined by the expression [17]

$$D = \left(1 + \frac{2m_e \omega L^2}{4\hbar}\right)^{-1} \tag{108}$$

Here, instead of the energy of a non-relativistic quantum particle, we use a Schrödinger frequency $\omega$ (which is equal to the difference between the true frequency of the electron wave that is entered into the solution of the Dirac equation and its "rest frequency" $\omega_e = mc^2/\hbar$ [10]). With the increase in the width of the potential step $L$, the transmission coefficient (108) decreases rapidly, and for an actual decelerating potential that has macroscopic sizes that substantially exceed the de Broglie wavelength $\lambda_{dB} = 2\pi\sqrt{\frac{\hbar}{2m_e\omega}}$, it is almost equal to zero because, in this case, we can neglect the tunnelling.

Thus, for the macroscopic decelerating potentials that are used in the experiments, there is a threshold effect: when $\hbar\omega > U_0$, the electron wave passes through the decelerating potential, while when $\hbar\omega \leq U_0$, the electron wave is fully "reflected" by the decelerating potential and the photoelectric current is not observed behind it. This arrangement means that there is a limit to the value of the decelerating potential, which is the stopping potential

$$U_s = \hbar\omega \tag{109}$$

above which the photoelectric current is absent.

When accounting for expression (107), we obtain

$$U_s = \hbar\omega_0 - \hbar|\omega_1| \tag{110}$$

This result completely coincides with the above given formulation of the second law of the photoelectric effect, and it was obtained within the framework of classical field theory without the use of such concepts as photons and electrons.

Note that expression (110) can be formally written in the form

$$\hbar\omega_0 = E + A \quad (111)$$

where the notations $A = \hbar|\omega_1|$ and $E = U_s$ were introduced. The expression in (111) can be considered to be Einstein's equation for the photoelectric effect, and one can interpret it within the framework of the photon-electron representations in which the parameter $E$ is interpreted as the kinetic energy of the photoelectrons, while the parameter $A$ is interpreted as a work function of the atom. However, this approach is no more than an interpretation that is based on the formal similarity of the pure wave expression (110) and the mechanical law of energy conservation.

The above analysis has shown that such a corpuscular interpretation of the photoelectric effect is superfluous.

The well-known experiments by E. Meyer and W. Gerlach (1914) on the photoelectric effect on the particles of metal dust, irradiated with ultraviolet light, are considered to be one of the pieces of "irrefutable evidence" that light energy is propagated in the form of identical indivisible quanta (photons). Assuming that the electrons are particles while light is composed of continuous classical electromagnetic waves, we can calculate the time during which the metal particle will absorb a sufficient amount of energy for the ejection of an electron. In the experiments by E. Meyer and W. Gerlach, this duration was of the order of a few seconds, which means that the photoelectron cannot leave a speck of dust earlier than in a few seconds after the start of irradiation. In contrast to this conclusion, the photoelectric current in these experiments began immediately after the beginning of the irradiation. Hence, it is usually concluded that this finding is only possible if the light is a flux of photons each of which can be absorbed by the atom only entirely and, therefore, can "knock out" the electron from the atoms at the moment of its collision with the metal particle.

However, this conclusion follows only in the case in which the electrons are considered to be indivisible particles. If instead of considering the electrons to be particles we consider a continuous electron wave [10,11], then as was shown above, the photoelectric current appears almost without delay after the start of irradiation of an atom by the classical electromagnetic wave and occurs even at very low light intensities, when the light frequency exceeds the threshold frequency for the given atom. This finding is because to start the photoelectric current, the atom does not need to accumulate the energy that is equal to the ionization potential because the electron wave is emitted by the atom continuously and not in the form of discrete portions – "electrons". Note that precisely the need to explain the ejection of discrete electrons from an atom under the action of light led A. Einstein to the idea of light quanta, which when absorbed, gave to the atom sufficient energy for the liberation of a whole electron.

The above analysis shows that all three laws of the photoelectric effect only approximately reflect its actual regularities. In particular, the photoelectric current appears and disappears non-abruptly when "passing" through the threshold frequency $|\omega_1|$, and it gradually increases or decreases in the frequency range that has the width $\Delta\omega_0 \sim \pi/t$ near the threshold frequency $|\omega_1|$. However, this effect can be detected only for ultrashort observation times of $t \sim 10^{-15}$ s, which is difficult to achieve in the experiments on the photoelectric effect. Moreover, consideration of the nonlinear effects in the interaction of the light wave with an atom shows [18] that the photoelectric current appears even in the case when the frequency of the incident light is significantly less than the threshold frequency $|\omega_1|$, which is predicted by the linear theory. Such effects can be observed only in a very intense laser field [19]. Strictly speaking, the theory [18], which describes the ionization of an atom in an intense laser field, is fully classical in the sense under consideration because an atom is described by the Schrödinger equation, while the light wave is considered to be a classical electromagnetic field. The true result of this finding is the photoelectric current that is created by the continuous electron wave emitted by an atom because precisely the photoelectric current is calculated in the theory [18]. However, traditionally, the results of the theory [18] are interpreted from the standpoint of photon-electron representations, which makes it necessary to interpret the main result of the theory [18] as the probability of the ionization of an atom (i.e., the probability of the liberation of an "electron" from the atom) per unit time. The representations with respect to the multiphoton ionization of an atom, when the atom "absorbs simultaneously several photons", the total energy of which exceeds the ionization potential of the atom, were a consequence of such an interpretation. When there is a requirement for too many "photons" for the liberation of the "electron", talking about the simultaneous absorption of such a large number of particles becomes meaningless (because of the low probability of this process); then, the results of the theory [18] are interpreted as a tunnel ionization in which the intense laser field changes the potential field in which the "electron" is positioned, which gives it the "opportunity" to leave the atom due to tunnelling. From the point of view of the ideas that are developed in this series of papers, both "multiphoton" and "tunnel" ionization of an atom are the result of the same process - the interaction of a classical electromagnetic wave with a classical electron wave.

Finally, note that there is no difficulty in calculating the angular distribution of the photoelectric current in the framework of the theory under consideration, if we account for the fact that the continuous electric current created by the electron wave emitted by an atom under the action of light is calculated by expression (88) using the wave function in (73) and (87). Once again, note that this current is not the distribution over the directions of the particles-electrons that are emitted by an atom, but the distribution over the directions of the current of a continuous charged electron wave that is emitted by the atom. All of the known expressions that are obtained earlier for the photoelectric effect (see, e.g., [14,16]) remain valid, but they should now be interpreted from the standpoint of classical field theory.

# 5  Concluding remarks

Thus, we see that the light-atom interaction is fully described within the framework of classical field theory without the use of quantum electrodynamics and, in general, without any quantization. The results of this theory utilize the simple classical sense and do not require the postulation of such paradoxical properties of matter as the wave-particle duality. The paradoxes in the theory appear when a continuous light beam or a continuous charged electron wave emitted by the atoms under the influence of incident light are attempted to be interpreted as the flux of indivisible particles - photons or electrons. In this case, the probabilistic interpretation of the results of the theory arises from a need. However, as was shown in this paper and in the previous papers of this series [10,11], the processes that are under consideration are fully deterministic, while the postulate about the probabilistic nature of all quantum phenomena is the result of misinterpretation.

In this paper, in the calculation of an atom-field interaction, we did not account for a property of an electron wave, such as the spin, i.e., the internal angular momentum and the associated internal magnetic moment of the electron wave that is continuously distributed in space [10]. To describe the atom-field interaction while accounting for the internal magnetic moment of the electron field, it is necessary to use the Dirac equation or, in the Schrödinger long-wave approximation [10], the Pauli equation, which should be supplemented by the terms that account for the inverse action of the self-radiation field on the electron wave. This issue will be considered in the subsequent papers of this series.

**Acknowledgments**

Funding was provided by Tomsk State University competitiveness improvement program.